\documentclass[11pt,twoside]{article}
\usepackage{asp2004}
\usepackage{psfig}
\usepackage{epsf}
\usepackage{graphics}
\usepackage{lscape}
\begin{document}
\thispagestyle{plain}
\title{Circular polarimetry of magnetic cataclysmic variables}

\author{Cl\'audia V. Rodrigues, Deonisio Cieslinski, Francisco J. Jablonski,
Fl\'avio D'Amico}
\affil{Instituto Nacional de Pesquisas
Espaciais --- Av. dos Astronautas, 1758 --- 12227-010 - S\~ao Jos\'e dos
Campos - SP --- Brazil}
\author{Jo\~ao E. Steiner, Marcos P. Diaz}
\affil{Inst. Astronomia, Geof\'\i sica e Ci\^encias Atmosf\'ericas/USP - Brazil}
\author{Gabriel R. Hickel}
\affil{IP\&D - Universidade do Vale do Para\'\i ba - Brazil}

\begin{abstract}
Magnetic cataclysmic variables are complex accreting binary systems with 
short orbital periods. Here we present circular
polarimetry of five magnetic cataclysmic variable candidates.
1RXS~J161008.0+035222, V1432~Aql, and 1RXS~J231603.9$-$052713 have
cyclotron emission, which confirms them as AM Her systems.
Our data are consistent with zero values for the circular polarization of
1RXS~J042555.8$-$194534 and FIRST~J102347.6+003841 imposing some constraints
to the polar classification of these objects.
\end{abstract}

\section{Introduction}

Cataclysmic variables (CV) are short period binaries composed by a white dwarf
(primary) and a late-type main-sequence star. The secondary star fills its Roche
lobe,
losing material to the primary by the inner Lagrangian point, L1. Due to its
angular momentum, this material forms an accretion disk around the white dwarf.
Polars (or AM Her systems) are CV in which the primary has a magnetic
field in the range from 10 to 100 MG. Such a high magnetic field
prevents disk formation. Therefore, the material from the secondary goes to the
primary magnetic pole(s) via an accretion stream. Another consequence of the
magnetic field is the synchronization of the white dwarf rotation with the
orbital revolution.

In the accretion region, near the white dwarf surface, the material is fully
ionized producing highly polarized cyclotron emission due to the presence of
the strong magnetic field. As a large fraction of
the optical flux in polars comes from such region, they have large
values of polarization, mainly circular. Consequently, polarimetry
is a fundamental tool to classify an object as an AM Her system. 

Stokes parameters are also a powerful accreting region diagnosis because they
depend strongly on the angle by which that region is observed as well
as on its physical properties. 
The orbital behavior of the linear polarization position angle depends
only on the inclination of the system, $i$, and the colatitude of the
axis of the magnetic field, $\beta$. Besides, the phase interval during which 
no cyclotron emission is observed also constrains $i$ and $\beta$.
Estimates of the value of the magnetic field as well as of some other plasma
properties can also be obtained through the modelling of flux and polarization
variation with orbital phase (see Wickramasinghe \& Meggitt 1985, for accreting
column cyclotron models).

In this work we present polarimetric observations of candidates to AM Her
systems for which no polarization measurement is reported in the literature.
A more detailed description of the observations and data reduction, as well as a
deeper analysis of the results, will be presented elsewhere. 
 
\section{Observations}

The observations have been done with the 1.6-m Perkin-Elmer
telescope at {\it Observat\'orio do Pico dos Dias}, operated by
the {\it Laborat\'orio Nacional de Astrof\'\i sica}, Brazil, using a
CCD camera modified by the polarimetric module described in Magalh\~aes et al.
(1996). The CCD arrays used were back-illuminated SITe with $1024 \times 1024$
pixels. The observations were done in three runs during 2003.

The polarimetric modulus consists of a fixed analyzer (calcite prism),
a $\lambda$/4 retarder waveplate and a filter wheel. 
The use of a calcite block, which separates the extraordinary and
ordinary beams, eliminates any sky
polarization (Piirola 1973; Magalh\~aes et al.  1996). The retarder plate
is rotated with 22\fdg5 steps. Therefore, a polarization measurement
consists of eight integrations in consecutive retarder
orientations. The $\lambda$/4
retarder allows us to measure the circular and linear polarization
simultaneously. 

The images have been reduced following the standard steps of differential
photometry using the IRAF\footnote{IRAF is distributed by
National Optical Astronomy Observatories, which is operated by the Association
of Universities for Research in Astronomy, Inc., under contract with the
National Science Foundation.} facility. Counts were used to
calculate the polarization using the method described in Magalh\~aes,
Benedetti, \& Roland (1984) and Rodrigues, Cieslinski, \& Steiner (1998). 
The polarimetric reduction was greatly facilitated by the
use of the package {\it pccdpack} (Pereyra 2000). Photometry can be done summing
the counts in the two beams.

\section{Results}

In the following sections, we present a short review of the objects and some
qualitative results.

\subsection{1RXS J042555.8$-$194534}

The ROSAT X-ray spectrum of 1RXS~J042555.8$-$194534 is typical of a polar
(Schwope
et al. 2000). Its optical spectrum and photometry are also consistent with this
classification (Schwope et al. 2002).
However, no circular polarization was detected at a level of 2\%. 
More observations - using time resolved optical spectroscopy, for instance - are
necessary to understand the nature of this object.


\subsection{FIRST J102347.6+003841}

\begin{figure}[!ht]
\plotfiddle{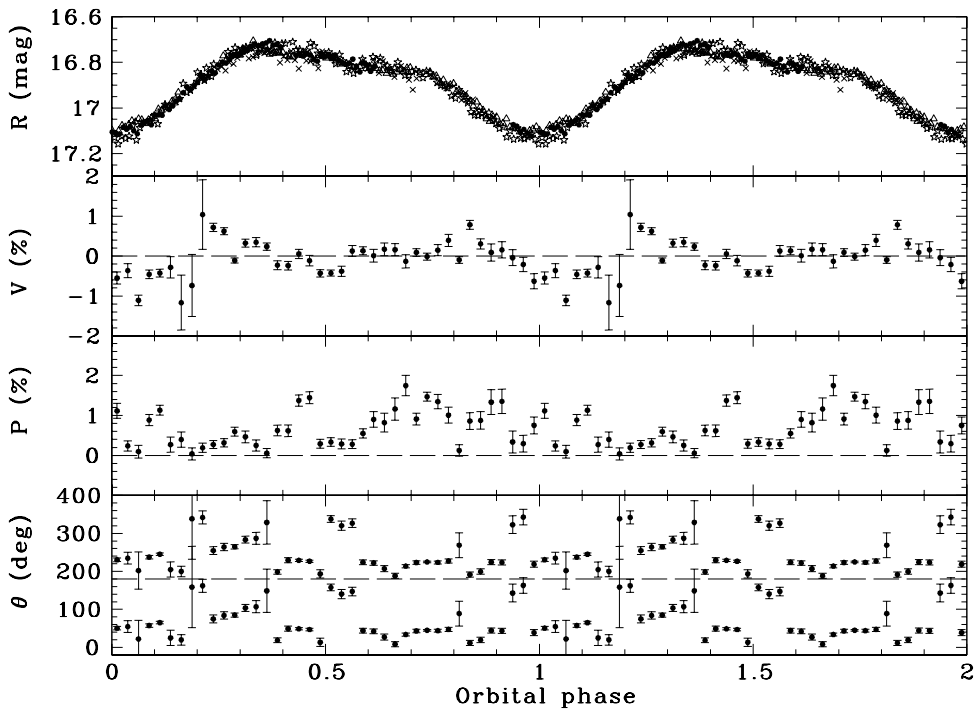}{9.0cm}{-90}{50}{50}{-190}{280}
\caption{Photometry, circular and linear polarimetry of 
FIRST~J102347.6+003841 in R band from data taken in April 2003.}
\label{first}
\end{figure}

This object was discovered by its radio emission and subsequent optical
observations revealed the spectrum of a magnetic CV (Bond et al. 2002).
Its optical light curve is compatible with a polar in which most of the flux
originates from the reflection of a hot white dwarf on the
secondary surface (Would,
Warner, \& Pretorius 2004). The circular polarization measured was
less than 1.5\% (Figure \ref{first}). This may be interpreted as a negligible
contribution from any cyclotron emission to the flux in the R band: less than
$\approx$ 1\%.

\subsection{1RXS J161008.0+035222}

\begin{figure}[!ht]
\plotfiddle{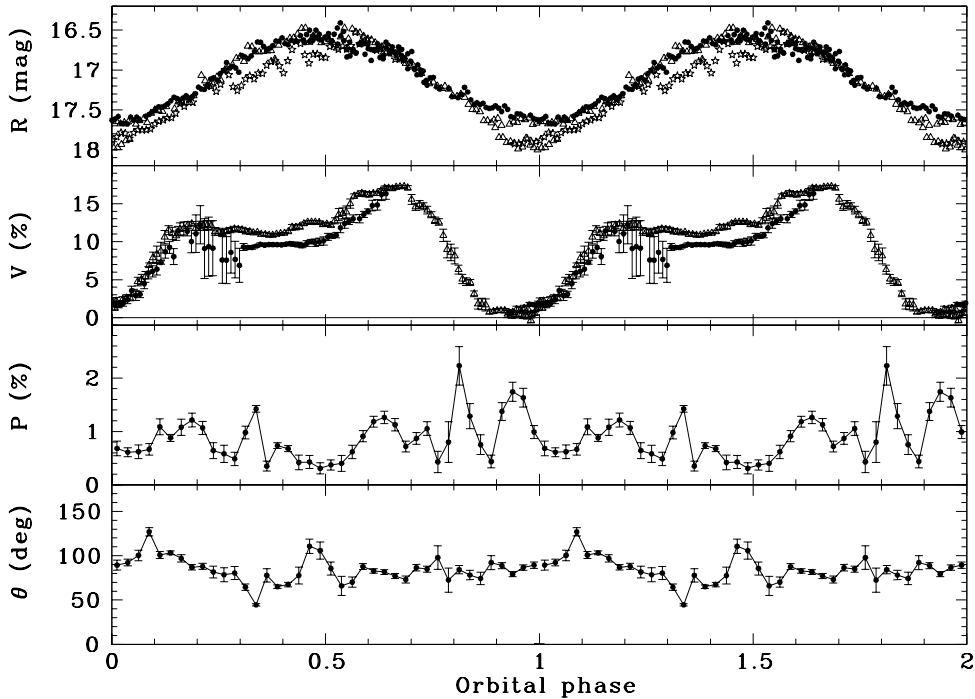}{8.5cm}{-90}{50}{50}{-190}{280}
\caption{Photometry, circular and linear polarimetry of
1RXS~J161008.0+035222 in R band from data taken in April 2003.}
\label{rx16}
\end{figure}

This source is another polar candidate from the ROSAT satellite (Schwope et al.
2000). Optical identification and spectral classification have been done by
Jiang et al. (2000) and Schwope et al. (2002). The high values of the measured
circular polarization (Figure \ref{rx16}) show that this object is a polar.
There was no evidence of occultation of the accretion region in the light
curve, indicating that the system may have an one-pole accretion geometry.
The small variation of the position angle is consistent with small values of
$\beta$. Another possible solution is 
$i \approx \beta \approx 90\deg$, but in this case  the accretion region
should be out of sight approximately half the orbital period.

\subsection{V1432 Aql}

In contrast with the other objects from this work, V1432 Aql is a well studied
CV.
Even so, its classification is still controversial:
an asynchronous polar  (Mukai et al. 2003) or
an intermediate polar (Singh \& Rana 2003)?
Previous attempts to detect circular polarization have only managed to put
upper limits to its value: 7\% (Watson et al. 1995) and 2\% (Friedrich et al.
1996).

We obtained measurements of V1432 Aql in two epochs: August and September 2003.
In both epochs, circular polarization of up to 4\% was measured.
If one considers the classification as an asynchronous polar and uses the
ephemerides for the orbital and white dwarf motions from Mukai et al. (2003),
the two epochs are separated by approximately half the beat period.
The polarization curves displayed different shapes in each observed epoch
even if phased with the white dwarf rotation.


\subsection{1RXS J231603.9$-$052713}

\begin{figure}[!ht]
\plotfiddle{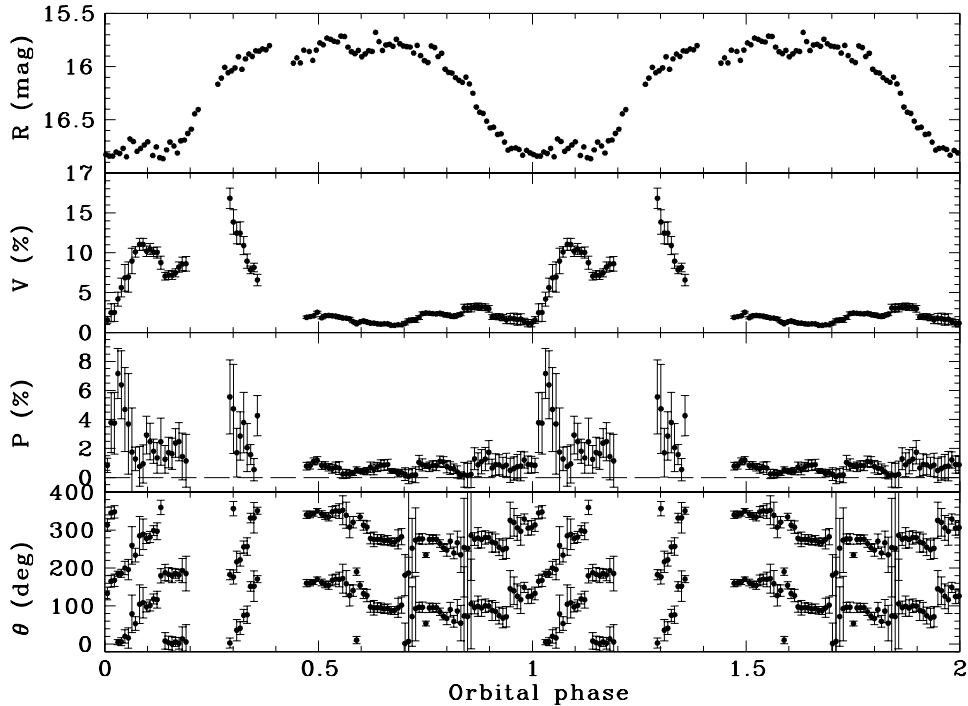}{8.5cm}{-90}{50}{50}{-190}{275}
\caption{Photometry, circular and linear polarimetry of
1RXS~J231603.9$-$052713 in R band from data taken in September~2003.}
\label{rx23}
\end{figure}

It is also a polar candidate from ROSAT (Beuermann \& Thomas 1993; Thomas et al.
1998; Schwope et al. 2000). In spite of the partial orbital period coverage, it
is evident a large circular polarization variable in phase confirming this
object as a polar (Figure \ref{rx23}).

\acknowledgements{CVR thanks the support from FAPESP (Proc. 03/13304-6 and 
01/12589-1).}

\end{document}